\documentclass[aps,prb,twocolumn,floatfix,showpacs,amsmath,amssymb]{revtex4}
\usepackage{graphicx}
\usepackage[hypertex]{hyperref}

\begin{document}

\newlength\smallfigwidth
\smallfigwidth=3.2 in
\newlength\figwidth
\figwidth=4.0 in

\preprint{UFV/UFJF/KSU Feb. 2005}

\title{Extinction of BKT transition by nonmagnetic
disorder in planar-symmetry spin models}

\author{G.\ M.\  Wysin}
\email{wysin@phys.ksu.edu}
\homepage{http://www.phys.ksu.edu/~wysin} \altaffiliation{
Permanent address: Department of Physics, Kansas State University,
Manhattan, KS 66506-2601 }\affiliation{ Departamento de F\'isica,
Universidade Federal de Vi\c cosa, Vi\c cosa, 36570-000, Minas
Gerais, Brazil }
\author{I.\ A.\ Marques}
\affiliation{ Departamento de F\'isica ICE, Universidade Federal
de Juiz de Fora, Juiz de Fora 36036-330, Minas Gerais, Brazil }
\author{S.\ A.\ Leonel}
\email{sidiney@fisica.ufjf.br}
\affiliation{
Departamento de F\'isica ICE,
Universidade Federal de Juiz de Fora,
Juiz de Fora 36036-330, Minas Gerais, Brazil
}
\author{P.\ Z.\ Coura}
\affiliation{ Departamento de F\'isica ICE, Universidade Federal
de Juiz de Fora, Juiz de Fora 36036-330, Minas Gerais, Brazil }
\author{A.\ R.\ Pereira}
\email{apereira@ufv.br} \affiliation{ Departamento de F\'isica,
Universidade Federal de Vi\c cosa, Vi\c cosa, 36570-000, Minas
Gerais, Brazil }

\date{March 14, 2005}

\begin{abstract}
The Berezinskii-Kosterlitz-Thouless (BKT) transition in
two-dimensional planar rotator and XY models on a square lattice,
diluted by randomly placed vacancies, is studied here using hybrid
Monte Carlo simulations that combine single spin flip, cluster and
over-relaxation techniques.
The transition temperature $T_c$ is determined as a function of vacancy density
$\rho_{\textrm{vac}}$ by calculations of the helicity modulus and the
by finite-size scaling of the in-plane magnetic susceptibility.
The results for $T_c$ are consistent with those from the much less precise
fourth-order cumulant of Binder.
$T_c$ is found to decrease monotonically with increasing $\rho_{\textrm{vac}}$,
and falls to zero close to the square lattice percolation limit,
$\rho_{\textrm{vac}}\approx 0.41$ .
The result is physically reasonable:  the long-range orientational order of
the low-temperature phase cannot be maintained in the absence of sufficient
spin interactions across the lattice.
\end{abstract}

\maketitle
\section{Introduction: Spin-Diluted Planar Spin Models}

It is well known that vortices are fundamental ingredients in the
Berezinskii-Kosterlitz-Thouless (BKT) phase transition.
\cite{Berezinskii70,Berezinskii72,Kosterlitz73} The simplest models
exhibiting this transition are the pure planar rotator model (PRM)
and XY-model. For these models the transition takes place at
critical temperatures $k_{B}T_{KT}/JS^2=0.898$ \cite{Tobochnik79}
and $k_{B}T_{KT}/JS^2=0.699$, \cite{Cuccoli+95,Evertz96}
respectively ($J$ is the exchange constant, $S$ the spin length).
Recently, the study of topological excitations such as vortices and
solitons in two-dimensional magnetic lattices containing defects has
received a lot of attention.
\cite{Subbaraman+98,Zaspel+96,Bogdan02,Pereira03,
Pereira+03,Wysin03,Paula+04,Pereira+05,Karatsuji96} Such
interactions must have interesting consequences for the static and
dynamical properties of easy-plane magnets. Analytical and numerical
calculations have shown that vortices are attracted and pinned by
nonmagnetic impurities. \cite{Pereira03,Pereira+03,Paula+04,Mol+03}
In fact, the vortex energy is lowered when pinned at a vacancy,
resulting in greater preference of single vortex \cite{Wysin03} and
vortex-pair \cite{Pereira04} formation on vacancies. Of course, this
leads to an overall increase of the system disorder.
All of these factors conspire to reduce the BKT transition
temperature with increasing vacancy density, as has already been
seen in calculations from Refs.\ \onlinecite{Leonel+03,
Berche+03,Wysin05} for planar spin models on a two-dimensional
square lattice (see also analytical results using the self-consistent
harmonic approximation of Ref.\ \onlinecite{Castro+02}).
The important question here is, at what vacancy density is the
transition temperature reduced to zero, so that the system is always
in the high-T disordered phase?
This would mean a situation in which there is no low-temperature phase
of long-range orientational order, characterized by spin-spin correlations
decaying as a power law with distance, and a finite absolute magnetization
$\langle | \sum_i \vec{S}_i | \rangle$ in the thermodynamic limit.
%

Calculations of the helicity modulus for the planar rotator model by
Leonel \textit{et al.}\cite{Leonel+03} indicated that a critical
vacancy density $\rho_{\textrm{vac}}=\rho_c \approx 0.3$ causes the
critical temperature $T_c$ to drop to zero. It means that the
critical temperature would vanish at $p_{c}=1-\rho_{c}\approx 0.7$,
which is above the site percolation threshold,
$p_{pt}=1-\rho_{pt}=0.59$, for a planar square lattice. Lozovick and
Pomirchi, \cite{Lozovik93} also using the jump in the helicity
modulus, have found that the BKT phase transition occurs above the
percolation threshold in a dilute system of Josephson junctions
(using bond dilution).
On the other hand, Berche \textit{et al.}\cite{Berche+03}
calculated the decay of the spin-spin correlation function and
its related exponent, $\eta$, and considered the transition
temperature to be located by $\eta(T_c) = 1/4$.
Those results suggested that the critical density is closer to
$0.41$ (the number associated with the percolation limit for the
square lattice).
The Monte Carlo calculations for this problem naturally are particularly
difficult, especially because the interesting region occurs at very low
temperature.
Furthermore, the statistical fluctuations due to the random choice
of positions for the vacancies further increases the numerical noise
in the calculations -- this effect itself becomes particularly
troublesome especially when $\rho_{\textrm{vac}}$ surpasses 0.3
(30\%).
As such, it seems important to make more reliable estimates for
the critical vacancy density based on improved MC calculations
here.

The calculations mentioned above concern the planar rotator
model (two-component spins lying in xy plane).
In a specialized model with repulsive vacancies, Wysin calculated the
reduction of $T_c$ in an easy-plane Heisenberg model, with three-component
spins with anisotropic couplings of their components.\cite{Wysin05}
The vacancies were not allowed to be on nearest or next nearest neighbor
lattice sites, which made it possible to calculate the vorticity
density in the model.
However, that calculation did not concern itself with the determination
of the critical vacancy density, because the constraint of repulsive
vacancies limits the possible vacancy density to be less than 18\%,
well below the critical value.
Therefore, for comparison with the planar rotator, we also
consider here the vacancy effects in the (three-component)
XY-model, with randomly placed non-repulsive vacancies.

After describing the model Hamiltonians, we give an overview of the
different methods used to estimate the transition temperature.
This is followed by some specific comments on the Monte Carlo schemes
applied to this problem.
The data obtained for the planar rotator and XY models are
presented, followed ultimately by our conclusions.

\begin{figure}
\includegraphics[angle=-90.0,width=\columnwidth]{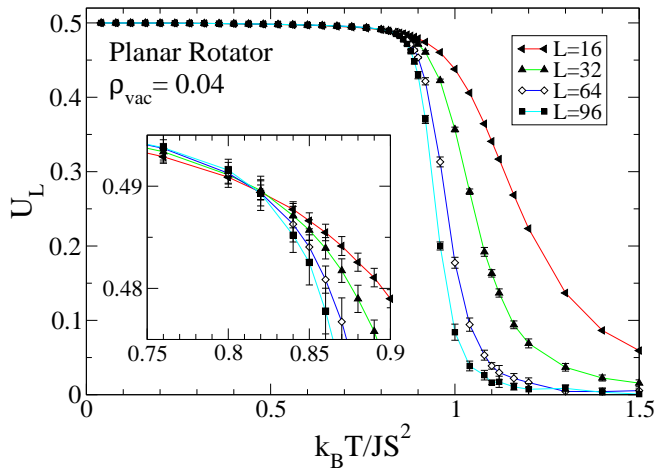}
\caption{
\label{U04_PR}
Application of the fourth order cumulant (\ref{UL}) for estimating $T_c$,
for the PRM at 4\% vacancy concentration.  The data were obtained using
the Monte Carlo approach described in Sec.\ \ref{MC-method}.  The inset
expands the view near the critical temperature ($k_B T_c/JS^2 \approx 0.815$).
}
\end{figure}

\section{Model Hamiltonians}
The spin models under consideration can be described by the Hamiltonian
\begin{equation}
H=-J\sum_{\langle i,j \rangle} \sigma_i \sigma_j
\left( S_i^x  S_j^x + S_i^y S_j^y \right) ,
\end{equation}
where $\langle i,j \rangle$ indicates nearest neighbor sites of an
$L \times L$ square lattice, and $J$ is the ferromagnetic exchange
coupling between spins $\vec{S}_i$ and $\vec{S}_j$.
The spins $\vec{S}_i$ have two components for the planar rotator
model and three components for the XY-model; in the latter case,
however, only the xy components are coupled.
The occupation variables $\sigma$ take the values $1$ or $0$ depending
on whether the associated site is occupied by a spin or vacant.
A fraction $\rho_{\textrm{vac}}$ of the sites are chosen randomly
to be vacant.
It is important to realize, however, that the Monte Carlo calculations
here must make adequate averages over different choices of the vacancy
positions, for a chosen density.

The planar rotator model has effectively a single degree of freedom
per site -- the angle of the spin within the xy plane.
The main distinction of the XY model is the presence of the
extra $S^z$ components, which act as degrees of freedom, but
do not appear in the Hamiltonian.
The XY model therefore involves two degrees of freedom per spin.
This increases the entropy effects at a given temperature and results
in a lower $T_c$ compared to the planar rotator.
The MC algorithm for the XY model must involve the possibility
to change all three spin components for the XY model,
while preserving the spin length.

\subsection{Physical properties leading to $T_c$}
The lack of significant sharp peaks in the thermodynamic
quantities versus temperature $T$ for these models, especially in
finite $L \times L$ lattice systems, means that precisely locating
$T_c$ is difficult.
Therefore, it is useful to apply several different approaches,
all essentially based on the scaling of the thermodynamics with
the system size or edge length $L$.

As the Monte Carlo algorithm proceeds, the total system
instantaneous in-plane magnetization $\vec{M}=(M_x,M_y)$  is
observed,
\begin{equation}
\vec{M} = \sum_i \sigma_i \vec{S}_i.
\end{equation}
Additionally, statistical fluctuations give the
susceptibility components for temperature $T$,
\begin{equation}
\label{chiaa}
\chi^{\alpha\alpha} = (\langle M_{\alpha}^2 \rangle
-  \langle M_{\alpha}\rangle^2)/(NT).
\end{equation}
The number of spins in the system is $N=(1-\rho_{\textrm{vac}})L^2$.
The average of $\chi^{xx}$ and $\chi^{yy}$ defines the in-plane
susceptibility,
\begin{equation}
\chi = \frac{1}{2}(\chi^{xx}+\chi^{yy}).
\end{equation}
%

A rough estimate of $T_c$ can be obtained from the size-dependence of
Binder's fourth order cumulant\cite{Binder81,Binder90}  $U_L$,  defined
by
\begin{equation}
\label{UL}
U_L = 1- \frac{ \langle (M_x^2 + M_y^2)^2 \rangle}
              {2 \langle M_x^2 + M_y^2 \rangle^2 }.
\end{equation}
For any $L$, the asymptotic values are $U_L(T\ll T_c)=0.5$, $U_L(T\gg T_c)=0$.
At the critical temperature, $U_L$ is approximately independent
of the system size, hence, $T_c$ can be estimated from the crossing
point of curves of $U_L(T)$ for various $L$.
An example of such crossing behavior is given in Fig.\ \ref{U04_PR}, for
the PRM at $\rho_{\textrm{vac}}=0.04$.
In practical application, due to the statistical uncertainties, there is
usually no clear crossing point, especially at higher vacancy concentrations.
Instead, $T_c$ is very close to the point where different
curves of $U_L(T)$ begin to separate from the low-T asymptotic value.
Although very reliable, this approach is not very accurate, and requires
MC calculations for many temperatures near $T_c$.

\begin{figure}
\includegraphics[angle=-90.0,width=\columnwidth]{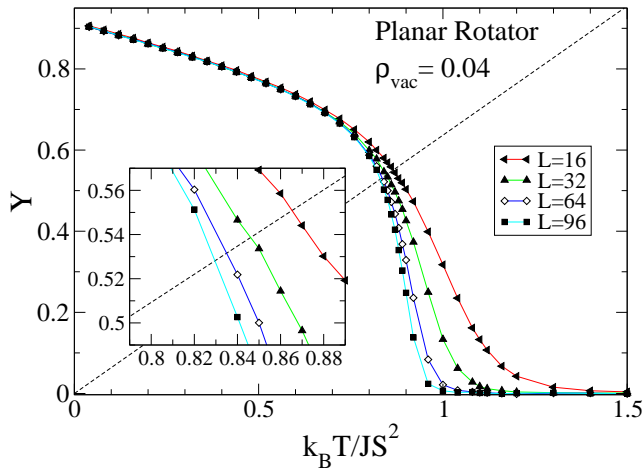}
\caption{ \label{Y04_PR} Typical application of the helicity modulus
for estimating $T_c$, for the PRM at 4\% vacancy concentration. The
dashed line is Eq.\ (\ref{line}).  The inset shows how the crossing
points occur slightly above the critical temperature ($k_B T_c/JS^2
\approx 0.815$).  Error bars are smaller than the symbols. }
\end{figure}

Another approach to determine $T_c$ is based on the calculation of the
helicity modulus per spin, $\Upsilon(T)$.
It is a measure of the resistance to an infinitesimal spin twist
$\Delta$ across the system along one coordinate, defined in terms
of the dimensionless free energy, $f=F/(JS^2)$,
\begin{equation}
\label{Y0}
\Upsilon = \frac{1}{N} \frac{\partial^2 f}{\partial \Delta^2}.
\end{equation}
Any general model Hamiltonian leads to the expression,
\begin{equation}
\label{Y1}
N\Upsilon = \left\langle \frac{\partial^2 H}{\partial \Delta^2} \right\rangle
-\beta \left[
   \left\langle \left(
     \frac{\partial H}{\partial \Delta}
                \right)^2 \right\rangle
- \left\langle
     \frac{\partial H}{\partial \Delta}
  \right\rangle^2
\right] ,
\end{equation}
where $\beta=(k_B T)^{-1}$ is the inverse temperature.
For either the planar rotator or XY model, the required operators
to be averaged (in limit $\Delta \to 0$) can be expressed
using the Cartesian spin components,
\begin{subequations}
\begin{equation}
G_s \equiv
\frac{\partial H}{\partial \Delta}
 =  \sum_{\langle i,j \rangle} \sigma_i \sigma_j
                  \left( \hat{e}_{i,j} \cdot \hat{x} \right)
                  \left(S_i^x S_j^y -S_i^y S_j^x \right) ,
\end{equation}
\begin{equation}
G_c \equiv
\frac{\partial^2 H}{\partial \Delta^2}
 =  \frac{1}{2} \sum_{\langle i,j \rangle} \sigma_i \sigma_j
                   \left(S_i^x S_j^x + S_i^y S_j^y \right) ,
\end{equation}
\end{subequations}
where $\hat{e}_{i,j}$ is a unit vector pointing from site $j$ to site $i$.
The sum determining $G_s$ only includes pairs of lattice sites displaced
by $\pm \hat{x}$.
Furthermore, one expects the mean of $G_s$ to be quite small, while
its fluctuations do contribute to the helicity formula (\ref{Y1}).
The sum for $G_c$ is seen to be proportional to the original Hamiltonian.

\begin{figure}
\includegraphics[angle=-90.0,width=\columnwidth]{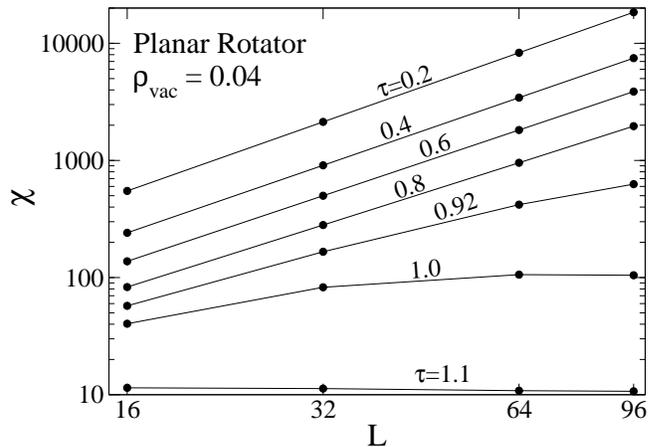}
\caption{ \label{Xfits04_PR} Log-log plot of susceptibility versus
system edge length $L$, for the PRM at 4\% vacancy concentration.
The curves correspond to different values of the dimensionless
temperature $\tau=k_B T/JS^2$.   Lines are guides to the eye; errors
are smaller than the symbols.  Least squares fits were used to
determine the slopes, $(2-\eta)$, producing $\eta(T)$ as seen in
Fig.\ \ref{eta04_PR}. }
\end{figure}

According to renormalization-group theory, the helicity modulus in an
infinite system jumps from the finite value $(2/\pi) k_B T_c$ to zero at the
critical temperature.
Assuming this applies also to the spin-diluted model, as argued in
Ref.\ \onlinecite{Leonel+03}, $T_c$ can be estimated from the
intersection of $\Upsilon(T)$ and the straight line,
\begin{equation}
\label{line}
\Upsilon = \frac{2}{\pi} k_B T.
\end{equation}
The trend in the intersection point with increasing $L$ can be observed,
as shown for the PRM at $\rho_{\textrm{vac}}=0.04$ in Fig.\ \ref{Y04_PR}.
Generally speaking, the MC data for $\Upsilon(T)$ show a steeper drop
in the critical region as $L$ increases.
The larger system size used, the lower will be the intersection point
and estimated $T_c$.
Hence this method will always lead to an over-estimate of $T_c$.

\begin{figure}
\includegraphics[angle=-90.0,width=\columnwidth]{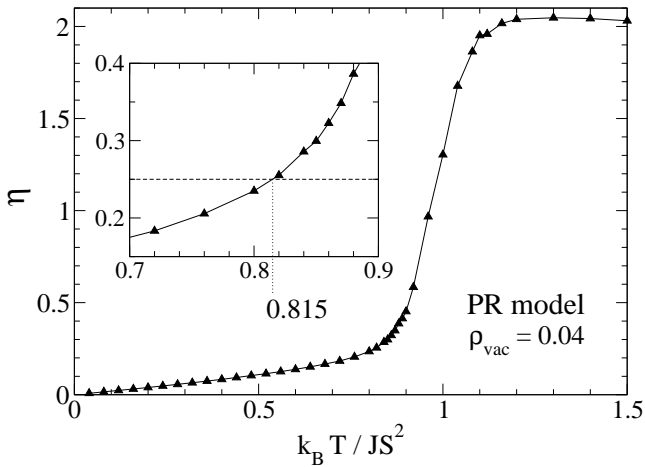}
\caption{ \label{eta04_PR} Application of the correlation exponent
$\eta$ for estimating $T_c$, for the PRM at 4\% vacancy
concentration, derived from using systems of sizes $L=16, 32, 64,
96$.  The inset shows how the critical temperature was estimated as
$k_B T_c/JS^2 \approx 0.815$. }
\end{figure}

Finally a third approach was also applied for estimating $T_c$,
employing the finite size scaling of the in-plane susceptibility, as
used in a pure XXZ model by Cuccoli \textit{et al.}\cite{Cuccoli+95}
and in the same model with repulsive vacancies by Wysin.
\cite{Wysin05}
In the absence of vacancies, it is the most precise method, because
the statistical fluctuations in $\chi$ can be reduced by extended MC
averaging much more effectively than those of the helicity modulus or
Binder's cumulant.
We assume that near and below $T_c$, a power law scaling of the susceptibility
holds, even in the presence of vacancies,
\begin{equation}
\chi \propto L^{2-\eta},
\end{equation}
where $\eta$ is the exponent for the in-plane spin correlations
below $T_c$ (see Ref.\ \onlinecite{Cuccoli+95}).
By using this equation with calculations at several lattice sizes,
the exponent $\eta$ can be fitted as a function of temperature.
An indication of how $\chi$ scales with system size is given in Fig.\
\ref{Xfits04_PR}, again for the PRM at 4\% vacancy concentration.
One can note clearly how the exponent $(2-\eta)$ (slope of log-log
plot for $\chi(L)$) decreases as the temperature increases, especially
rapidly as $T$ passes the transition temperature.

For the pure PR and XY models (no vacancies),  the transition is located
at the temperature where $\eta(T)=1/4$.
Then, under that assumption that the vacancies do not change the
basic symmetries in the transition, but only increase the
effective entropy present, we can expect that the transition can
be located in the same way under the presence of vacancies,
solving
\begin{equation}
\label{DefTc}
\eta(T_c) = \frac{1}{4}.
\end{equation}
In the absence of any particular theory for the model with vacancies,
this can be expected to be a reasonable definition for $T_c$.
Analysis of power-law fits of the spin-spin correlations in the
diluted PRM \cite{Berche+03} also demonstrated that $T_c$ occurs
very close to the temperature from Eq.\ (\ref{DefTc}).
Its validity is further verified here by the comparison with the
results for $T_c$ due to the helicity modulus, and due to Binder's
cumulant, the latter of which is reliable for any kind of model,
with or without vacancies.
Fig.\ \ref{eta04_PR} shows its application for the PRM at 4 \%
vacancy concentration, leading to $k_B T_c/JS^2 \approx 0.815$,
exactly consistent with the estimate from Binder's cumulant (Fig.\
\ref{U04_PR}).

We also note, that for the pure PRM (no vacancies), this fitting of
$\eta$, using systems as large as $L=160$, leads to the estimate
$k_B T_c/JS^2 =0.907 \pm 0.004$, slightly higher than that from
Ref.\ \onlinecite{Tobochnik79}.

\begin{figure}
\includegraphics[angle=-90.0,width=\columnwidth]{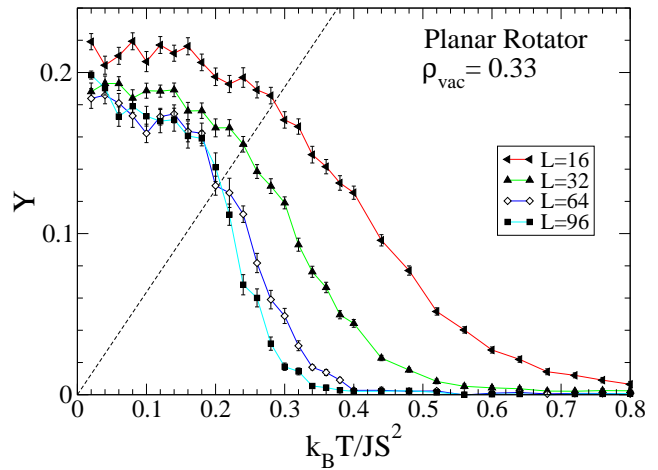}
\caption{ \label{Y33_PR} The helicity modulus for the PRM at 33\%
vacancy concentration for system sizes indicated.  The dashed line
is Eq.\ (\ref{line}). }
\end{figure}

\begin{figure}
\includegraphics[angle=-90.0,width=\columnwidth]{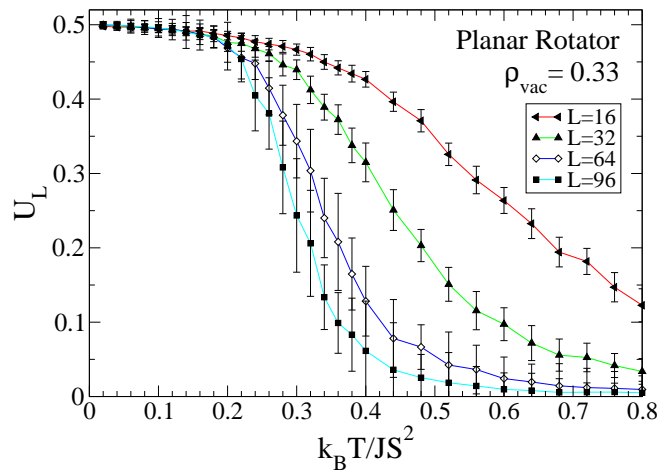}
\caption{ \label{U33_PR} Binder's fourth cumulant for the PRM at
33\% vacancy concentration for system sizes indicated.  $k_B T_c
/JS^2 \approx 0.14$ as estimated from the point where the data for
different system sizes separate. The lines are guides to the eye. }
\end{figure}

\subsection{Monte Carlo Scheme}
\label{MC-method}
Thermal averages for a given system size and temperature were obtained
using a hybrid MC approach, including Metropolis single-spin moves and
over-relaxation moves\cite{Evertz96} that can modify all spin
components, in combination with Wolff single-cluster
moves\cite{Wolff88,Wolff89} that modify only the xy components.
These are based on standard approaches for spin models, as developed in
many references.\cite{Kawabata+86a,Kawabata+86b,Wysin90,Gouvea97,Landau99}
Further details, as applied to a system with vacancies,
can be found in Ref.\ \onlinecite{Wysin05}.
Using a hybrid method including cluster and over-relaxation is very
important especially at low temperatures, as it very effectively reduces
the problems associated with critical slowing down.

The programming used for the XY model also applies to the planar rotator
model;  it is only necessary to set the out-of-plane components
$S^z=0$ and then never allow them to change for the PRM.
Thus it is straightforward to study the two models with essentially
the same MC approach.

The calculations were performed for a range of system sizes,
including $L=16, 32, 64, 96,$ and $160$.
For a given $L\times L$ lattice, the number of vacancies placed at random
locations was $N_{\textrm{vac}}=\rho_{\textrm{vac}}L^2$ (spins
removed from system or equivalently, set to zero length).
For larger systems or very low vacancy density, the results are nearly
independent of the particular random choice of vacancy positions.
In the general case, however, it is necessary to average over equivalent
systems (same $L$, $\rho_{\textrm{vac}}$) with different particular
choices of the vacancy locations.
The statistical variations in the averages are most significant especially
as the vacancy density approaches the critical value that forces $T_c$
to zero.
These statistical variations also are strongest in the smaller systems;
conversely, the larger systems tend to average out these fluctuations,
all the better as their area increases.
Therefore we averaged over a number $N_{\textrm{sys}}$ copies of the system,
with this number taken largest at small system size.
For $\rho_{\textrm{vac}} < 0.35$,  we used $N_{\textrm{sys}}=64, 32, 8, 4,$
for $L=16, 32, 64, 96$, respectively.
For larger density, $\rho_{\textrm{vac}} > 0.35 $, we doubled these values
for $N_{\textrm{sys}}$, and additionally included runs with
$N_{\textrm{sys}}=4$ for $L=160$.

\begin{figure}
\includegraphics[angle=-90.0,width=\columnwidth]{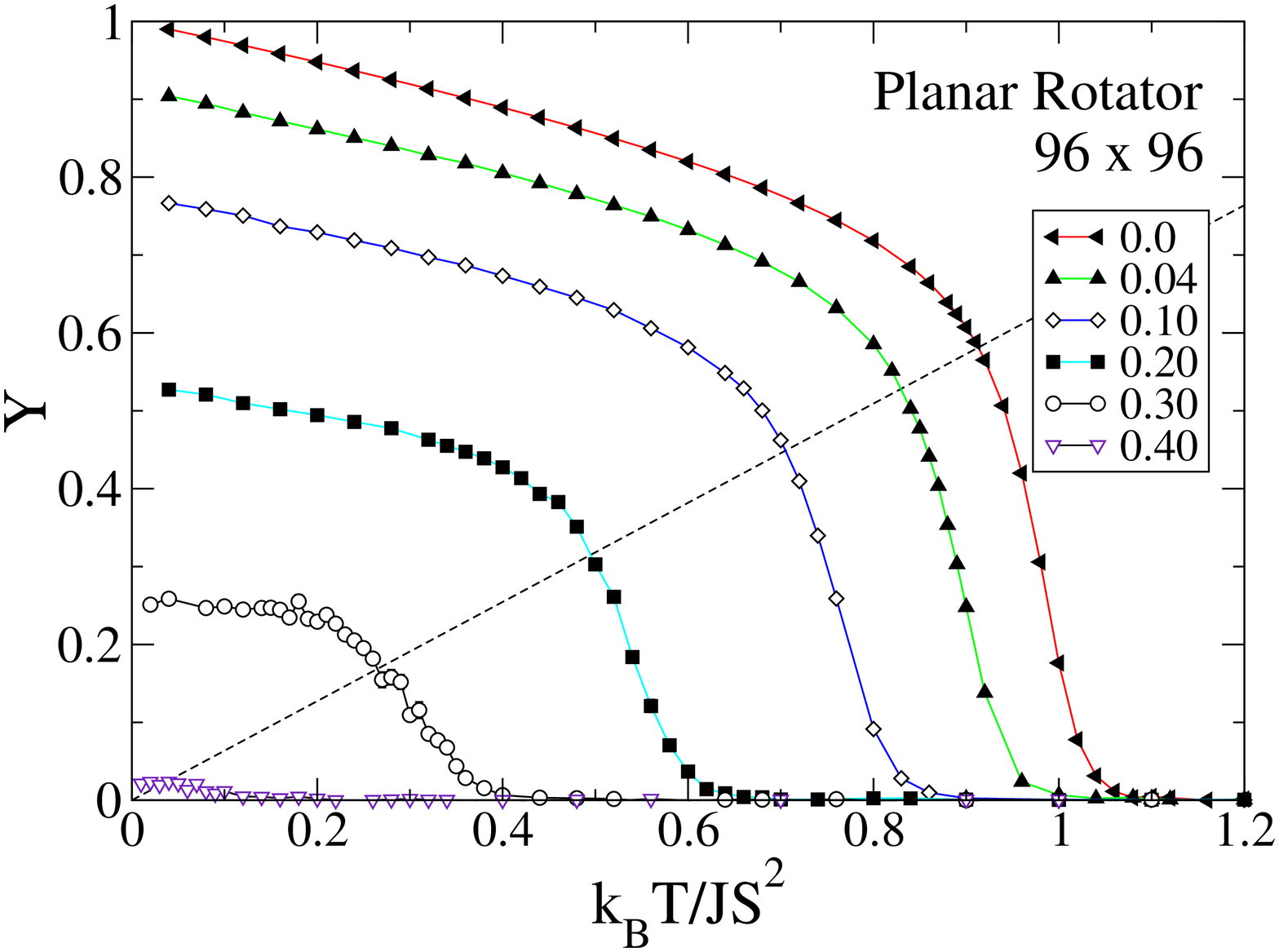}
\includegraphics[angle=-90.0,width=\columnwidth]{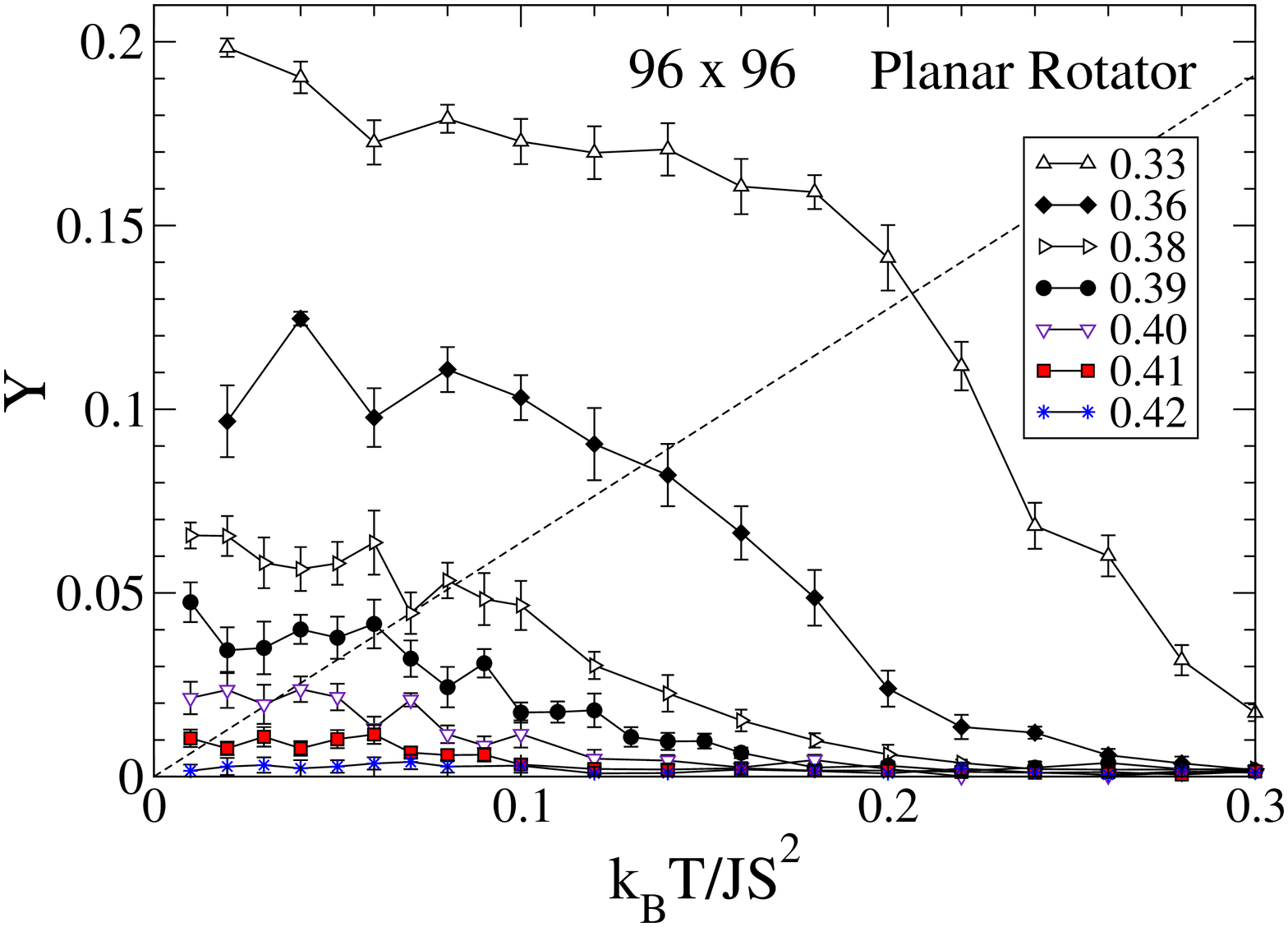}
\caption{ \label{Y_PR} The helicity modulus for the PRM at vacancy
concentrations $\rho_{\textrm{vac}}$ indicated in the legend.  The
dashed line is Eq.\ (\ref{line}).  Part (a) shows the overall trend;
error bars are smaller than the symbols.  Part (b) displays the
behavior as the transition is extinguished at the critical vacancy
concentration. }
\end{figure}

For thermal equilibration before calculating averages, $5000$ MC steps
(MCS)\footnote{One MC step includes \textit{attempting} one single spin
move per site and one over-relaxation move per site, and individual
cluster generation until the total number of sites in all clusters
formed has reached $(1/4)L^2$ sites.}
were applied for small systems ($L < 50$) and
$10,000$ MCS for large systems.
For each of the $N_{\textrm{sys}}$ individual realizations of a given $L$ and
$\rho_{\textrm{vac}}$,  averages at one temperature were calculated using
between 20,000 and 80,000 MCS ($N_{\textrm{data}}$), with the greatest number
applied to the larger systems.
For example, calculation for one temperature of a $16 \times 16$ lattice
at $\rho_{\textrm{vac}}=0.1$ involved an average over
$64 \times 25,000 = 1.28 $ million MCS.
On the other hand, one temperature of a $96 \times 96$ lattice at
$\rho_{\textrm{vac}}=0.36$ involved an average over
$8 \times 80,000 = 640,000$ MCS.
Near 0\% vacancy density, these MC parameters produce insignificant
error bars;  when $\rho_{\textrm{vac}}$ exceeds 30\%, on the other hand,
the resulting error bars are considerably greater and resist reduction.
As suggested above, the error bars in $\Upsilon$, $\chi$, and $U_L$
can be reduced more readily by increasing $N_{\textrm{sys}}$ than by
increasing $N_{\textrm{data}}$ when significant vacancy density is
present (especially at $\rho_{\textrm{vac}}>0.3$).

\begin{figure}
\includegraphics[angle=-90.0,width=\columnwidth]{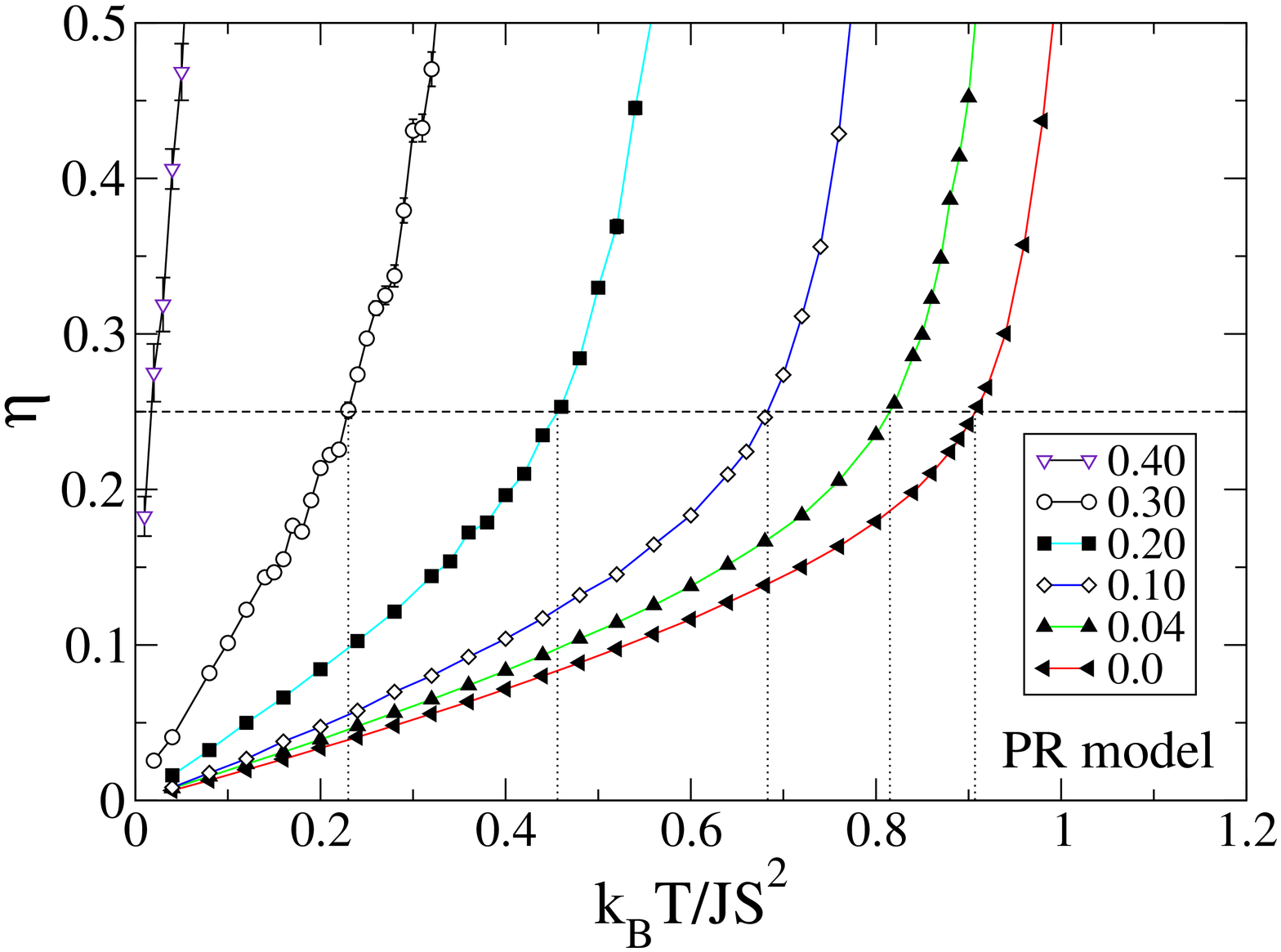}
\includegraphics[angle=-90.0,width=\columnwidth]{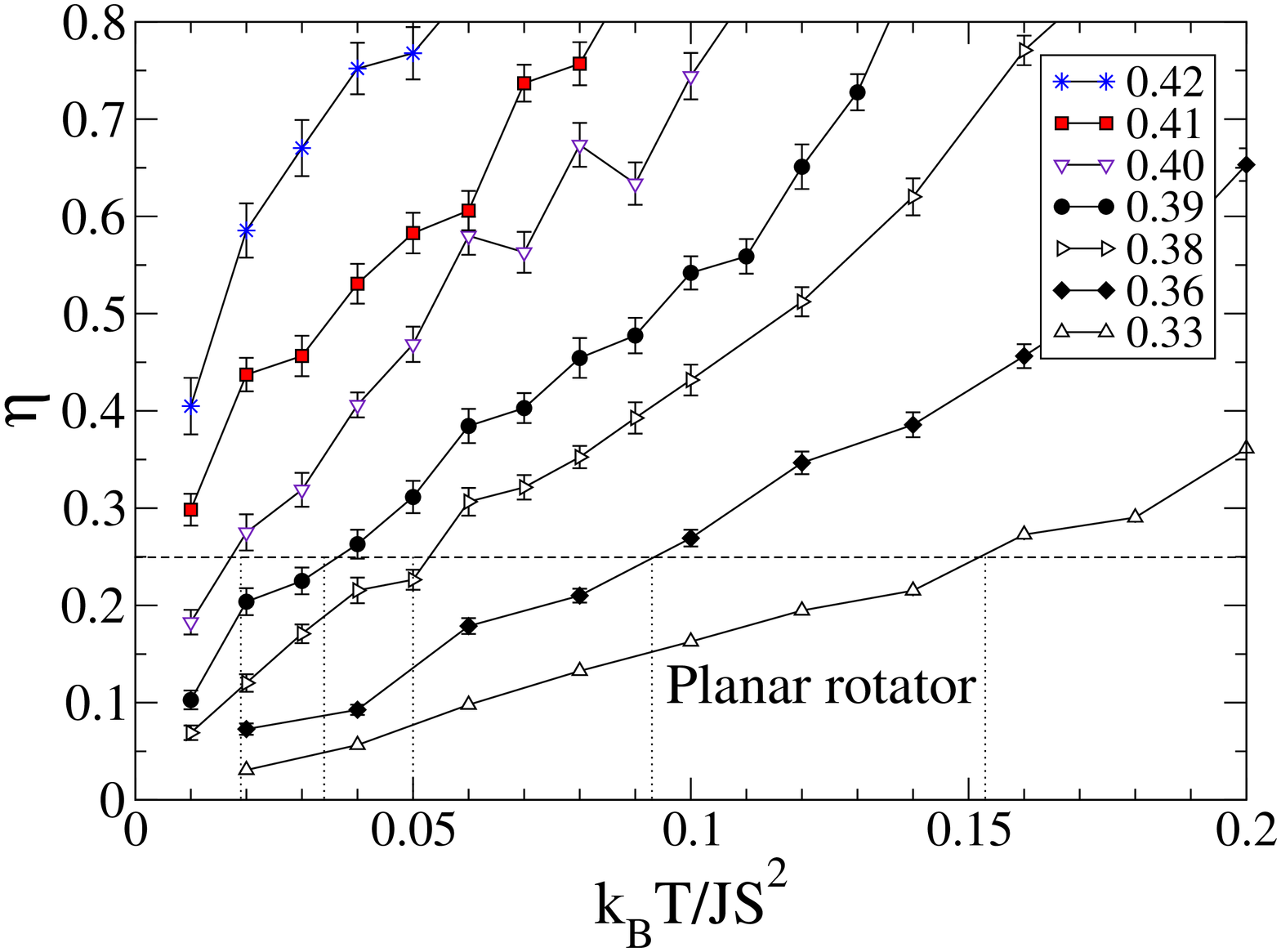}
\caption{ \label{eta_PR} Application of the correlation exponent
$\eta$ for estimating $T_c$, for the PRM at the vacancy
concentrations indicated in the legend, derived from scaling of
$\chi$ for systems of sizes $L=16, 32, 64, 96$. Part (a) gives the
rough overall trend and part (b) shows how $\eta$ does not fall to
the value $1/4$ at vacancy concentrations greater than 41\%. }
\end{figure}

\section{Monte Carlo Data}
Calculations were carried for a range of vacancy densities from
zero to 50\%.
We especially concentrated on the region
$0.30 < \rho_{\textrm{vac}} < 0.40$, which required the most
careful analysis.
For vacancy density less than 30\%, it is clear that there is
a transition at a finite temperature, for both the PR and XY models.
At the higher vacancy concentrations, statistical errors were generally
more significant.
Even so, looking at the trends in the data with system size, in the
following we show the MC evidence that the transition temperature
is reduced to zero when the vacancy concentration is approximately
40\%, for both models.

\subsection{Planar rotator model}
At low vacancy concentrations ($\rho_{\textrm{vac}}<0.20$), MC
results for $U_L$, $\Upsilon$, $\chi$, and $\eta(T)$ bear a great
resemblance to those shown above for 4\% vacancies, with fairly
smooth dependencies on temperature.
The primary modification is the general trend of important features
towards lower temperature with increasing $\rho_{\textrm{vac}}$.
At higher concentrations, errors become more significant.
For example, the helicity modulus at 33\% vacancies and various system
sizes is shown in Fig.\  \ref{Y33_PR}.
In addition to larger relative errors, the absolute magnitude of
$\Upsilon$ is drastically reduced.
It is very clear, however, that the BKT transition is still present
at this concentration, with $k_B T_c/ JS^2 \approx 0.20$ as estimated
from the crossing point of the $L=96$ data.
This is additionally supported by the corresponding behavior of
Binder's cumulant, seen in Fig.\ \ref{U33_PR}, which gives the
estimate $k_B T_c/ JS^2 \approx 0.14$, somewhat lower, as can be
expected.

An indication of the tendency for reduction of $T_c$ with vacancy
concentration is given in Fig.\ \ref{Y_PR}, showing a collection
of results all for $L=96$ systems.
While these crossing points consistently overestimate $T_c$,
a better view of this critical point reduction is provided by
the various graphs of $\eta(T)$ at different concentrations,
Fig.\ \ref{eta_PR}.
One can see clearly that once the vacancy concentration passes
a value around 41\%,  the fitted value of $\eta$ does not fall
below the value $1/4$, at least for the lowest temperatures
used ($k_B T/JS^2 =0.01$).
%

\begin{figure}
\includegraphics[angle=-90.0,width=\columnwidth]{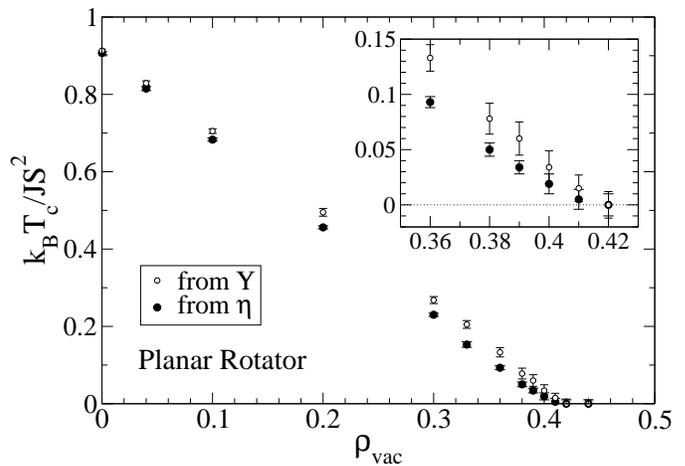}
\caption{ \label{Tc_PR} The critical temperatures versus vacancy
concentration for the PRM, extracted from fits of $\eta$ together
with Eq.\ (\ref{DefTc}) and from the crossing of $\Upsilon(T)$ with
Eq.\ (\ref{line}) for $L=96$ systems.  The inset shows $T_c$ as
$\rho_{\textrm{vac}}$ approaches the critical region. }
\end{figure}

In Fig.\ \ref{Tc_PR},  the critical temperatures extracted from $\eta$
and Eq.\ (\ref{DefTc}) and from the helicity modulus (using $L=96$) are
shown as functions of vacancy concentration.
As mentioned above, the crossing point of $\Upsilon(T)$ with Eq.\
(\ref{line}) for any finite system always overestimates $T_c$, with
less error as larger systems are used.
The fitting of $\eta$ is more reliable; the data shown here used the
scaling fitting of $\chi$ using systems with sizes $L=16, 32, 64, 96$.
The numerical values of $T_c$ estimated using $\eta(T_c)=1/4$ are
summarized in Table \ref{TcTable}.
The results give strong evidence for extinction of the BKT transition
at a vacancy concentration close to 41\%.

\begin{table}
\caption{\label{TcTable} Dependence of dimensionless critical temperature
$\tau_c \equiv k_B T_c/ JS^2$ on $\rho_{\textrm{vac}}$, as derived from the
scaling of in-plane susceptibility $\chi$, and using $\eta(T_c)=1/4$.
}
\begin{ruledtabular}
\begin{tabular}{cccc}
$\rho_{\textrm{vac}}$  &  $\tau$ (PR model)  & $\tau$ (XY model)\\
\hline
0.0  &  $0.907 \pm 0.004$  &   $0.700 \pm 0.005$ \\
0.04 &  $0.815 \pm 0.005$  &   $0.637 \pm 0.005$ \\
0.10 &  $0.683 \pm 0.004$  &   $0.547 \pm 0.005$ \\
0.16 &       ---           &   $0.453 \pm 0.005$ \\
0.20 &  $0.456 \pm 0.004$  &   $0.384 \pm 0.005$ \\
0.30 &  $0.230 \pm 0.004$  &   $0.208 \pm 0.005$ \\
0.33 &  $0.153 \pm 0.007$  &   $0.147 \pm 0.005$ \\
0.36 &  $0.093 \pm 0.005$  &   $0.087 \pm 0.005$ \\
0.38 &  $0.050 \pm 0.006$  &   $0.049 \pm 0.005$ \\
0.39 &  $0.034 \pm 0.006$  &   $0.041 \pm 0.005$ \\
0.40 &  $0.019 \pm 0.009$  &   $0.018 \pm 0.007$ \\
0.41 &  $0.005 \pm 0.009$  &   $0.003 \pm 0.007$ \\
0.42 &  $0.0   \pm 0.005$  &   $0.0   \pm 0.005$ \\
0.44 &  $0.0   \pm 0.005$  &   $0.0   \pm 0.005$ \\
\end{tabular}
\end{ruledtabular}
\end{table}

\subsection{XY model}
The general trends in MC data for the XY model are rather similar
to those found for the planar rotator.
The most obvious distinction, however, is that the extra entropy
due to the out-of-plane spin component forces the transition
temperature to be lower in the XY model, no matter what vacancy
concentration is considered.

\begin{figure}
\includegraphics[angle=-90.0,width=\columnwidth]{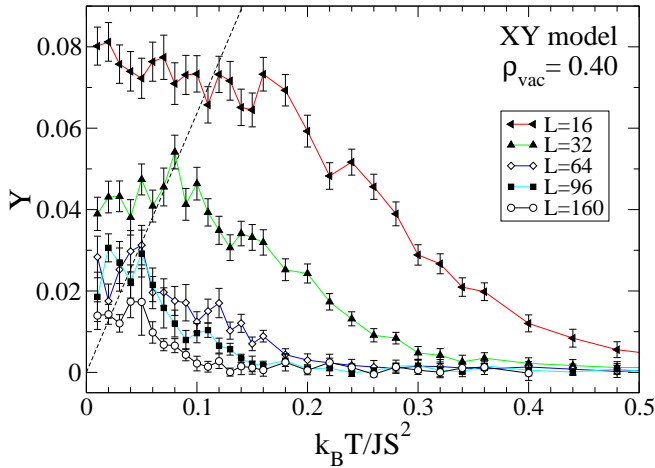}
\caption{
\label{Y40_XY}
The helicity modulus for the XY model at 40\% vacancy concentration
for system sizes indicated.  The dashed line is Eq.\ (\ref{line}).
}
\end{figure}

\begin{figure}
\includegraphics[angle=-90.0,width=\columnwidth]{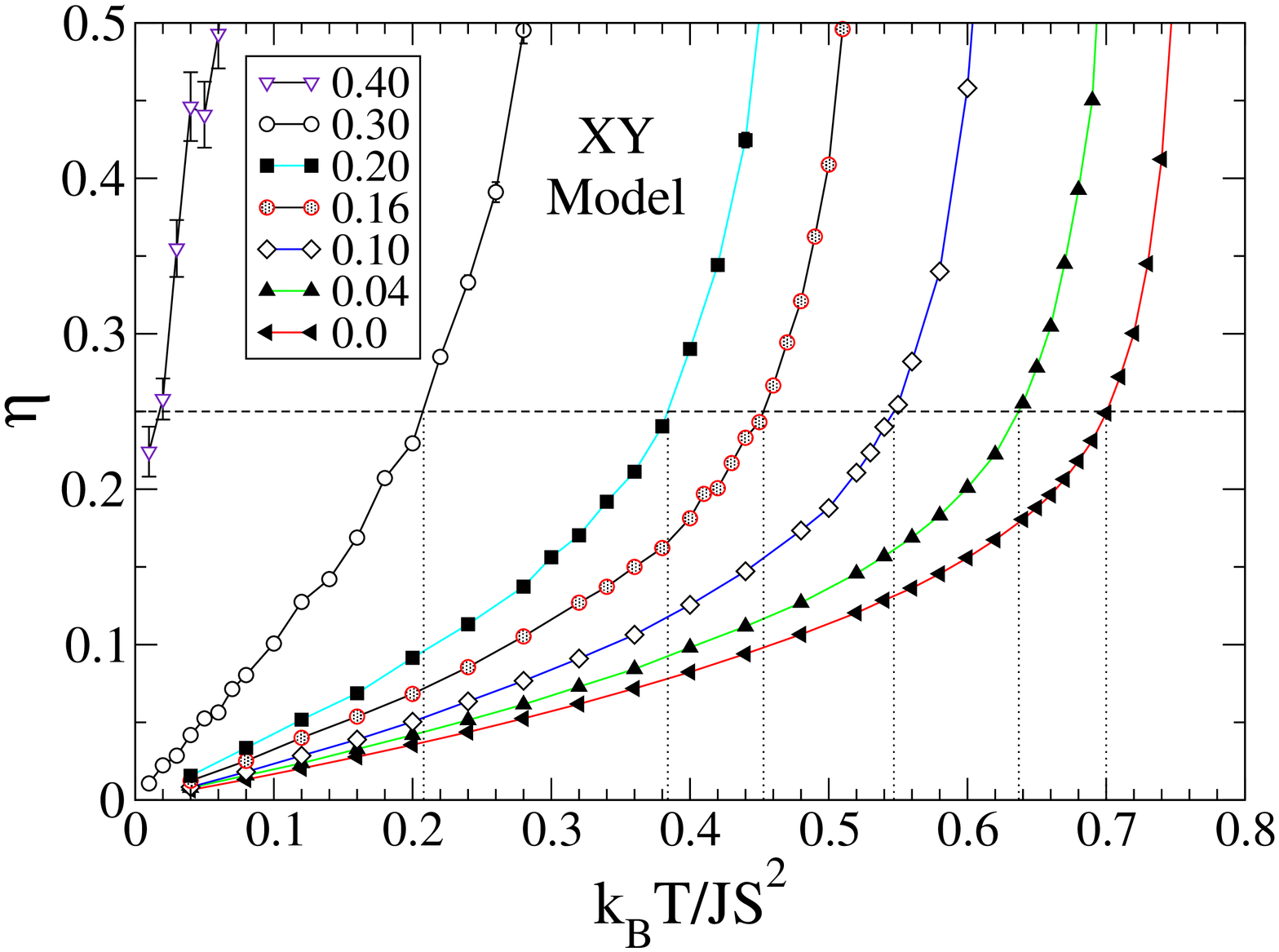}
\includegraphics[angle=-90.0,width=\columnwidth]{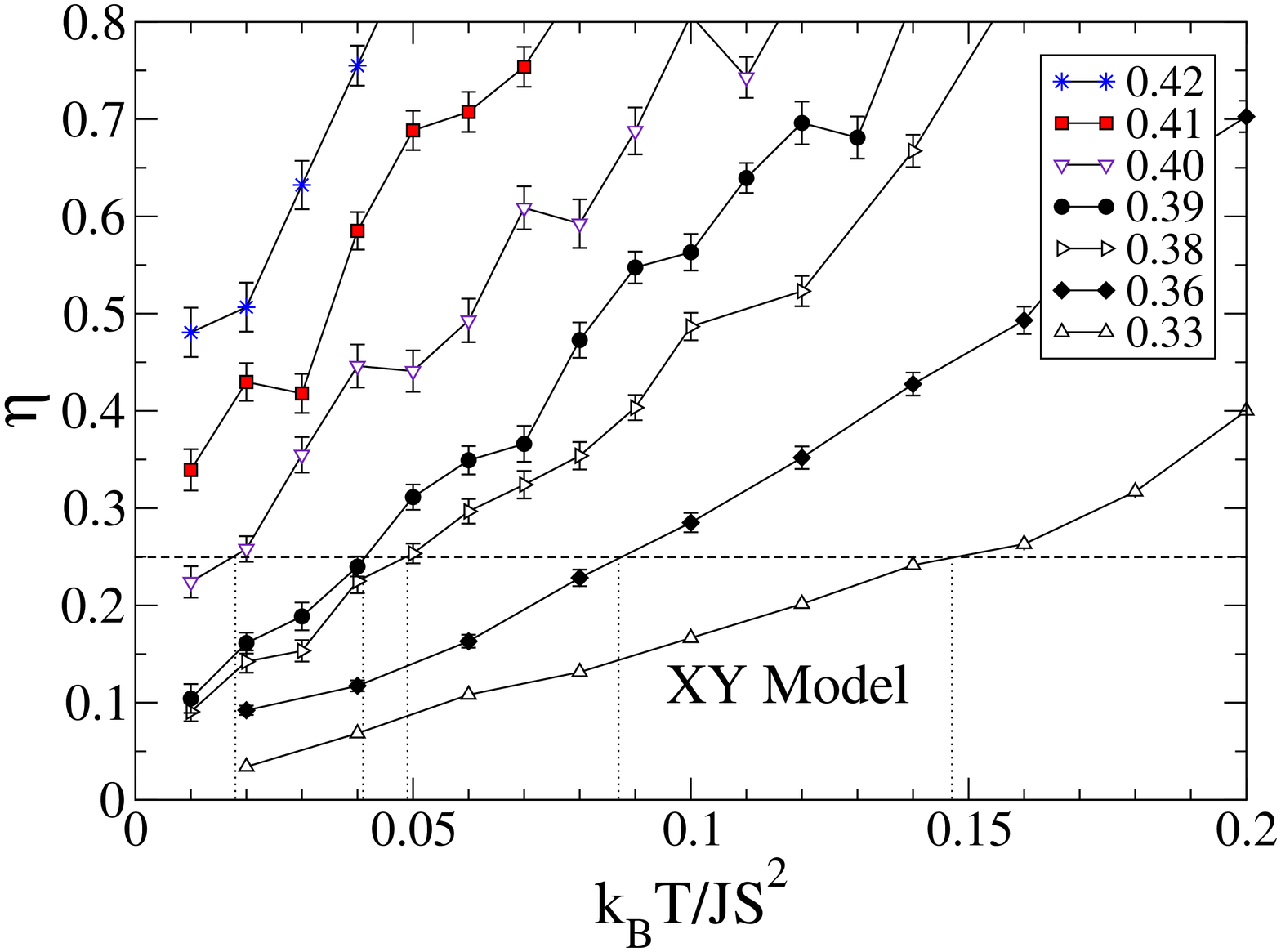}
\caption{
\label{eta_XY}
Application of the correlation exponent $\eta$ for estimating $T_c$,
for the XY model at the vacancy concentrations indicated in the
legend, derived from scaling of $\chi$ for systems of sizes
$L=16, 32, 64, 96$.
Part (a) gives the rough overall trend and part (b) shows how $\eta$
does not fall to the value $1/4$ at vacancy concentrations greater
than 41\%.
}
\end{figure}

It is interesting to show some data at 40\% vacancy concentration,
where the transition is seen to occur very slightly above zero
temperature.
In Fig.\ \ref{Y40_XY} the helicity modulus for system sizes from
$L=16$ to $L=160$ is displayed.
As the data for increasing system size is seen to systematically
fall to lower values, this graph alone cannot undeniably prove the
presence of a transition.
However, when taken in conjunction with the fits for $\eta$, which
passes the value $1/4$ around $k_B T/JS^2 \approx 0.018$, we
can say that even at 40\% vacancy density there occurs a
transition at finite temperature.
This can be seen in Fig.\ \ref{eta_XY}, where $\eta(T)$ is shown
for the various vacancy concentrations studied.
On the other hand, performing the MC calculations at temperatures
as low as $k_B T/JS^2 =0.01$, the exponent $\eta$ does not acquire
such a low value as $1/4$ even for 41\% vacancy concentration.

\begin{figure}
\includegraphics[angle=-90.0,width=\columnwidth]{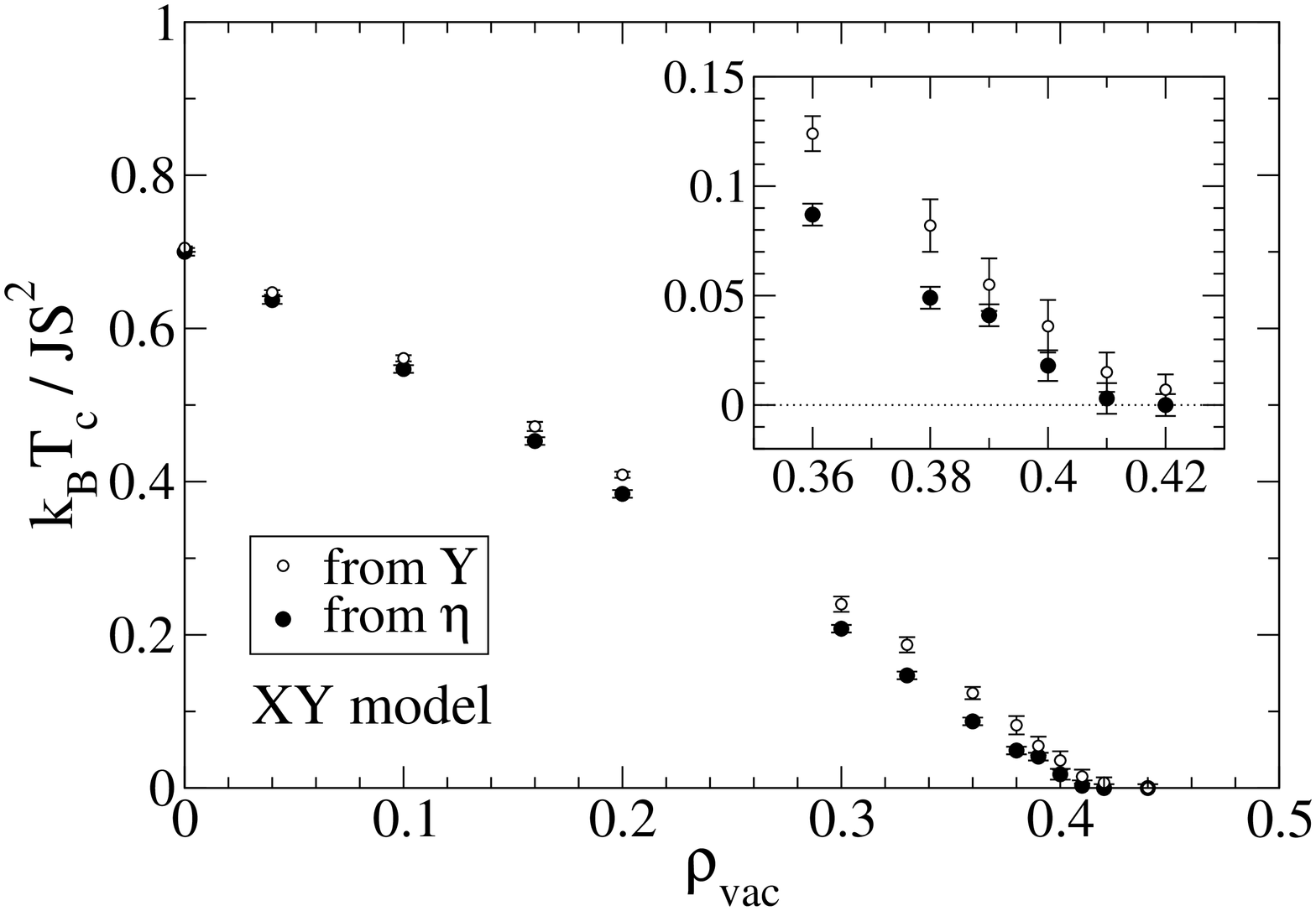}
\caption{
\label{Tc_XY}
The critical temperatures versus vacancy concentration for the
XY model, extracted from fits of $\eta$ together with Eq.\ (\ref{DefTc})
and from the crossing of $\Upsilon(T)$ with Eq.\ (\ref{line})
for $L=96$ systems.  The inset shows $T_c$ as
$\rho_{\textrm{vac}}$ approches the critical region.
}
\end{figure}

In Fig.\ \ref{Tc_XY},  the critical temperatures extracted from $\eta$
and Eq.\ (\ref{DefTc}) and from the helicity modulus (using $L=96$) are
shown as functions of vacancy concentration.
The numerical values as derived using $\eta(T_c)=1/4$ are given in
Table \ref{TcTable}.
Just as in the PR model, these results demonstrate the extinction of
the BKT transition at a vacancy concentration close to 41\%.
As the transition is controlled by the in-plane spin components,
the presence of the extra $S^z$ component in the XY model changes the
overall scale of transition temperatures, but does not affect the
critical vacancy concentration.


\begin{figure}
\includegraphics[angle=-90.0,width=\columnwidth]{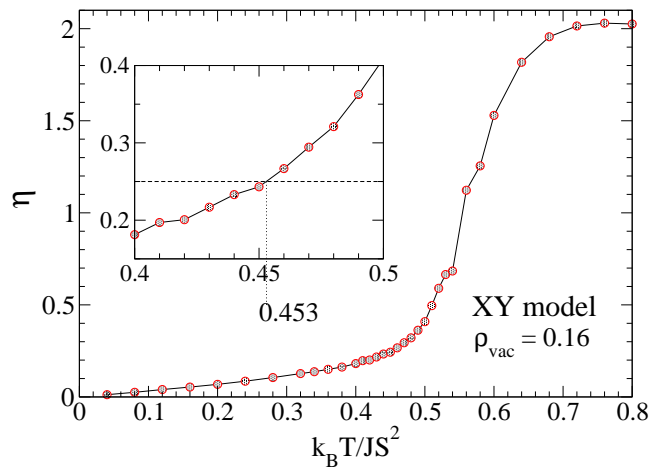}
\caption{
\label{eta16_XY}
Application of the correlation exponent $\eta$ for estimating $T_c$,
for the XY model at 16\% vacancy concentration, derived from using
systems of sizes $L=16, 32, 64, 96$.  The inset shows how the critical
temperature was estimated as $k_B T_c/JS^2 \approx 0.453$.
}
\end{figure}

Finally, we can also make some comparison to the XY model using
repulsive vacancies studied in Ref.\ \onlinecite{Wysin05}.
Considerable data was presented there for the case of 16\% vacancies.
Therefore it is interesting to note how the transition temperature is
changed if the vacancies are allowed to be at completely random positions
in the current model.
%

\begin{figure}
\includegraphics[angle=-90.0,width=\columnwidth]{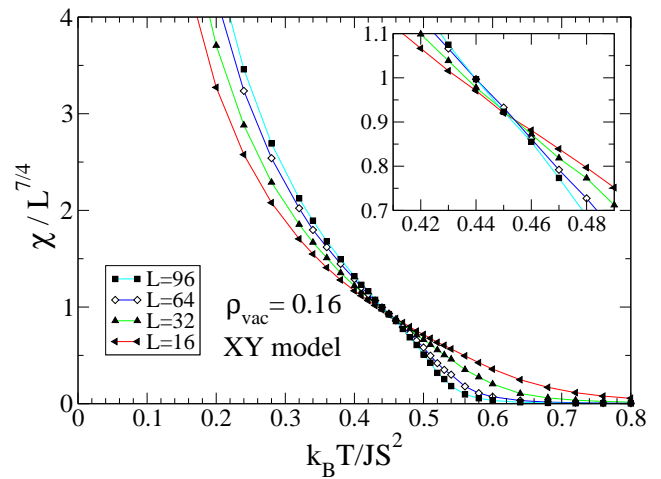}
\caption{
\label{Xs16_XY}
Application of the finite-size scaling of in-plane susceptibility $\chi$
to estimate $k_B T_c/J S^2 \approx 0.453$ (common crossing point of the data)
at 16\% vacancy density in the XY model, with fully randomly placed
vacancies, using exponent $\eta=1/4$.
}
\end{figure}

A graph of $\eta(T)$ for this case is given in Fig.\ \ref{eta16_XY},
showing clearly the transition occurring at $k_B T_c/JS^2 \approx
0.453$.
Alternatively, and even with less computational effort, the transition
can be found as done in Refs.\ \onlinecite{Cuccoli+95,Wysin05} by
plotting $\chi/L^{(2-\eta)}$ vs. $T$, taking $\eta=1/4$, and looking
for the common crossing point of data at various system sizes.
This is seen in Fig.\ \ref{Xs16_XY}, which gives the same estimate
for $T_c$.
In the repulsive vacancy model at the same vacancy concentration,
the transition occurs at a slightly higher temperature,
$k_B T_c/JS^2 \approx 0.478 \pm 0.001$.
The result is reasonable; there is greater disorder in the model
with fully random vacancies, hence, requiring less thermal
disordering due to temperature to reach the high-temperature phase.
(alternatively, the repulsive vacancy model has more built-in order
and hence requires greater thermal energy per spin to reach the
high-temperature phase.)

\section{Conclusions}
Hybrid MC calculations applied to the planar rotator and XY models
on a 2D square lattice show that the BKT transition is extinguished
($T_c \to 0$) at a vacancy concentration close to 41\%, a number
related to the percolation limit.
Then, although the BKT phase transition has an unusual nature, in
which the topological long-range order is destroyed by the unbinding
of vortices, the percolation problem of systems exhibiting such a
transition must have some similarities to the traditional $2D$ Ising
model.
In general, the transition temperatures for the XY model are
lower than those for the PR model, due to the extra entropy of
out-of-plane spin motions, but otherwise, the static properties
are closely related.
The transition temperatures were determined most precisely using
the finite-size scaling of the in-plane magnetic susceptibility,
under the assumption that the spin-correlation exponent $\eta$ goes
to the universal value $1/4$ at the transition, regardless of the
vacancy concentration.
This is equivalent to saying that the presence of spin vacancies
does not change any fundamental symmetries of the problem.
$T_c$ calculated this way is completely consistent with the
corresponding results from the helicity modulus and Binder's fourth
order cumulant.
At vacancy concentration higher than 41\%, the intrinsic disorder
of the system always produces a phase with short range correlations
that decay exponentially, i.e., the usual ``high-temperature''
BKT phase whose properties are strongly determined by the presence
of unbound vortices and antivortices.
The lack of percolation across the system at
$\rho_{\textrm{vac}}>0.41$ disrupts the ability to generate
topological long-range correlations.
It then becomes impossible to lower the temperature adequately to
reach the ordered phase of very low vortex density, dominated by
spin waves.

\begin{acknowledgments}
GMW is very grateful for support from FAPEMIG as a visiting
researcher and for the hospitality of the Universidade Federal de
Vi\c cosa, Brazil. ARP thanks CNPq for financial support.
\end{acknowledgments}

\end{document}